# Multidirectional analysis of the oscillating (T = 24 hours) Earth's electric field recorded on ground surface.

Thanassoulas[1], C., Klentos[2], V., Verveniotis, G.[3]

1. Retired from the Institute for Geology and Mineral Exploration (IGME), Geophysical Department, Athens, Greece.
   e-mail: thandin@otenet.gr - URL: www.earthquakeprediction.gr

2. Athens Water Supply & Sewerage Company (EYDAP),
   e-mail: klenvas@mycosmos.gr - URL: www.earthquakeprediction.gr

3. Ass. Director, Physics Teacher at 2nd Senior High School of Pyrgos, Greece.
   e-mail: verveniotis_ge@hotmail.com - URL: www.earthquakeprediction.gr

**Abstract**

The Earth's preseismic oscillating (T = 24h) electric field recorded for a short-time period of some days is analyzed in terms of its intensity vector azimuthal direction calculated at one monitoring site. The calculated azimuthal directions are compared to the concurrent seismicity observed for the same period of time. Examples are presented for proving the agreement between the electric field intensity vectors calculated azimuths and the corresponding ones referring to the EQs – monitoring site location. Finally, an example is presented on account of the use of this methodology upon three different monitoring sites for the utilization of the estimation of the epicentral area of a large future EQ at the Methoni, Greece seismogenic area.

**1. Introduction**

The evolution of seismic prediction was, at its start, initiated with efforts to determine the time of occurrence of the next strong EQ, which would take place in a seismogenic area. These efforts, intrinsically, incorporated the assumption that, the seismogenic area was known in advance. Therefore, this approach represents the first approximation of the epicentral area determination of a pending, strong EQ. This notion explains the reason, why a lot of efforts were dedicated in the time prediction, at the early stages of the evolution of earthquake prediction.

Another prognostic parameter, that is the location of the future, strong EQ, needed a statistical treatment of the past seismic history of regional areas, in order to reach a conclusion about the probability for the occurrence of a strong EQ in a predefined, seismogenic area. Such types of studies are referred, as "spatial-temporal" studies. In these studies, the spatial distribution of strong EQs and the time of their occurrence are interrelated, so that some well-defined rules are justified, answering both questions "where" and "when" the next strong EQ will take place. Some typical examples, of such studies, were presented by: Wyss and Baer (1981), for the case of Earthquake Hazard of the Hellenic Arc, Keilis-Borok and Rotwain (1990), when they studied the diagnosis of time of increased probability for strong earthquakes in different regions of the world, using the algorithm CN, Romachkova et al. (1998), when they performed intermediate-term predictions in Italy, using the Algorithm M8, Sobolev (2001), when he studied the earthquake preparation in Kamchatka and Japan, using the RTL parameter, Sobolev et al. (2002), when they studied the phases of earthquake's preparation and by chance test of seismic quiescence anomaly in conjunction to the RTL method for the Kamchatka region, Di Giovambattista and Tyupkin (2004), when they studied the seismicity patterns before the M=5.8, 2002, in Palermo (Italy), earthquake along to seismic quiescence and accelerating seismicity, Ogata and Zhuang (2006), when they analyzed Space-time ETAS models.

An, entirely, different problem is posed when the seismological data of an area, are very poor or non-existent. It is obvious that in such a case, spatial-temporal methodologies will fail. This was the case of Kozani, Greece EQ (M=6.6R, 1995) and Athens, Greece EQ (M=5.9R, 1999). Both areas were considered as "safe", in terms of seismic hazard, but Nature proved unpredictable. Therefore, it is evident that, it is necessary a method to exist which will provide, in a way, the epicentral area, independently from the past statistical, seismicity study of the regional area.

A logical question to be asked is: where is this specific earthquake going to take place. Solving this problem, independently from seismic statistics and previous knowledge of seismogenic areas, it requires the activation of a mechanism, which is capable to modify the physical properties of the regional space in a way that the modified physical parameters exhibit directional properties. In other words it is possible to measure directional properties of a generated field which will "point towards" the epicentral area of a future EQ.

Mizutani et al. (1976), after having studied electrokinetic's phenomena, associated, with earthquakes, suggested "if we are capable of detecting the electric current induced by the water flow, we can know the direction of the water flow, and consequently predict the epicenter of the earthquake".

Varotsos et al. (1981, 1984), related the observed, telluric, precursory signals intensity to the epicentral distance of an earthquake from different, registering, monitoring sites. The determination of the epicentral area was utilized by the use of the "Apollonian circles" and the "**1/r**" law, representing the signal intensity, as a function of distance from the seismic epicenter.

In a closer in terms of physics, different approach, Thanassoulas (1991, 1991a), using as a basis recordings of the VAN group, proved that it is possible to determine, the epicentral area of a pending strong EQ, much more accurately, by triangulating the calculated, azimuthal directions of the observed, electric earthquake precursory signals at different monitoring sites. The philosophy behind this methodology is that: although, the intensity of the observed, anomalous field depends on the local, geological conditions, the azimuthal direction of its intensity does not change, since it depends mainly on the regional current flow or the static field distribution, in the crust.

Ifantis et al. (1993) used the very same methodology for the determination of the epicenters of two strong EQs in Greece.

Pham et al. (2001) conducted a small-scale, field experiment at CRG, Garchy France. They recorded, signals in various sites originating from leak currents from the building complex. The origin of these currents was successfully determined by analyzing the azimuthal direction of the recorded signal at each monitoring station.

The three last works, which are referred, starting with Thanassoulas (1991) to Pham et al. (2001) clearly, indicate, that the epicentral area of an impending, strong earthquake can be determined, very accurately, by considering azimuthal directions calculation and triangulation of the anomalous, observed, electrical field, instead of any other statistical method used, so far.

An entirely different methodology, for determining the epicentral area of a strong EQ, is based upon ionospheric perturbations, observed, over the epicentral area, a few days before the seismic event takes place, due to ground tectonic, related, physical processes. The oldest related literature which is traced in various publications is the one by Moore (1964) and Davies and Baker (1965) which deals with the strong EQ of Good Friday, in Alaska, in 1964. More recent papers, related, to the topic of epicenter determination by the use of the study of the ionospheric perturbation, were presented by Depueva and Rotanova (2001), Pulinets et al. (2003), Pulinets (2004), Pulinets et al. (2004), Pulinets (2006).

The problem for the determination of the epicentral area of a strong EQ is generalized as follows: is it possible to identify the location of a future strong EQ, in a wide region, of unknown seismic activity in the past, and of no other geological or tectonic information, available?

Although, as a first approach, the answer to this problem appears to be impossible or rather very difficult, it will be shown that, a deterministic solution exists, based on basic, physics principles of electrical, potential fields. The interesting feature of any electrical field is the fact that it exhibits directional properties. In other words, it is possible to calculate the origin of its generating mechanism, by measuring its components at different locations and by inverting them, after having adopted an appropriate, generating, spatial, physical model.

The first and most basic assumption is: **the crust – lithosphere system, behaves as a homogeneous media,** as far as it concerns its electrical properties for large wavelengths fields, generated in it (Thanassoulas, 1991). This assumption drastically, facilitates the utilization of further mathematical operations which are required for the determination of the epicentral area. A strong objection, which could be raised, immediately, is: "but the crust – lithospheric system is not homogeneous, in terms of present geology and tectonics". Well, this is true, if short wave lengths of electrical fields are used compared, with lateral, geological – tectonic discontinuities. On the contrary, if large wavelengths of electric field are used, then the effect of geology – tectonics does not exist and the crust – lithospheric system becomes "transparent" for such electrical field wavelengths and consequently, it behaves as homogeneous media. Moreover, if the basic, adopted, assumption is false, then the results which are obtained, by using a "false" methodology, will be "false", too. In case the methodology, to be presented, provides "true" results, complying with the location of actually seismic, strong events which have occurred, then it proves that the basic, adopted assumption is valid.

Various different valid, electrical fields – signals generating mechanisms, have already been referred, in the scientific literature, (Thanassoulas, 2007). Each scientist defends his own seismic, signal generating mechanism by using robust arguments. The real problem, posed by the earthquake prediction requirement, is: which generating mechanism is valid at each seismic event case, so that it could be used, if possible, for epicenter area determination.

Moreover, what is the physical model, which could be used for calculations of this kind? Further more, if more than one, still unknown, physical mechanisms, is, preseismically, triggered, is it possible to "invert" the resulted, combined field, into its "origin" location? It is evident that, as long as the generating mechanisms which are triggered before any strong earthquake, are unknown, the more difficult, not to consider it as impossible, the solution of this problem, is.

A solution for this problem, based on principles of Applied Geophysics, was presented by Thanassoulas (1991). In this case, the geophysical notion of "apparent", physical value was applied to the case of the seismogenic area which generates electrical, preseismic signals, due to various, triggered, physical mechanisms. In particular, **the total, observed, electrical field that results from the combination of the different sub-electrical fields which are generated by the various, triggered, physical mechanisms, is attributed to a single, fictitious, point current source (Apparent Point Current Source). This current source generates, exactly, the very same total preseismic, electrical field.**

This is the second assumption upon which the entire methodology is based on. Its validity will be proved by the results which are obtained by applying the methodology on real, preseismic, electrical data.

Schematically, the entire notion of the apparent point current source is presented in the following figure **(1)**.

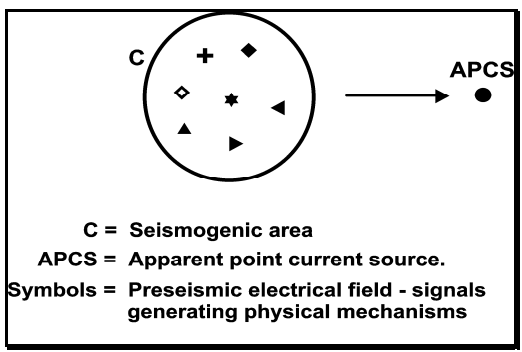

Fig.1. Various physical mechanisms (symbols), generating electrical, preseismic signals, triggered, in a seismogenic area **(C),** substituted by a fictitious "apparent" point current source **(APCS)** which generates the very same total preseismic electrical signals.

The main advantage of the adoption of the **APCS** is the simplification of the mathematical analysis, as far as it concerns the determination of the epicentral area of a strong future earthquake.

A generalized, representative model, for the case of generation of a preseismic, electrical field, is presented in the following figure **(2)**.

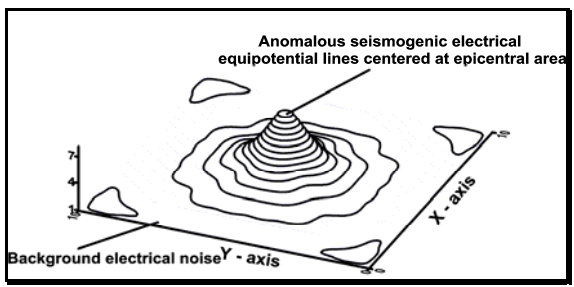

Fig. 2. The preseismic, electrical field, which is generated, by the **APCS** forms circular, equipotential lines on ground surface, centered at the epicentral area, is presented. The horizontal plane represents, apart from the ground surface, the ambient noise level, present, in the seismogenic area.

The preseismic, electrical signal (field), in most favorable cases, when a strong EQ is pending, exceeds the ambient noise level and thus is easily detectable. In cases, when the signal (field) amplitude is of lower level than the ambient noise, then it requires specific methodologies to be applied, so that the signal to noise ratio is improved at an acceptable and useful level.

This generalized model, of the preseismic, electrical field, which is observed over the focal area, provides a physical explanation for the ionospheric perturbations (electron, plasma densities), which are observed over the epicentral areas, some days prior to the occurrence of strong EQs. The assumed mechanism indicates that the generated, preseismic electrical field penetrates (Horn et al. 2007) the ionospheric layers at heights of almost 100-200Km and therefore, acting as a static, electric field lens, it modifies, circularly, the spatial distribution of the plasma ions, which align along the equipotential surfaces of the preseismic, electrical field which is present, at these heights.

The perturbing mechanism is shown, in a simplified presentation, in the following figure **(3)**.

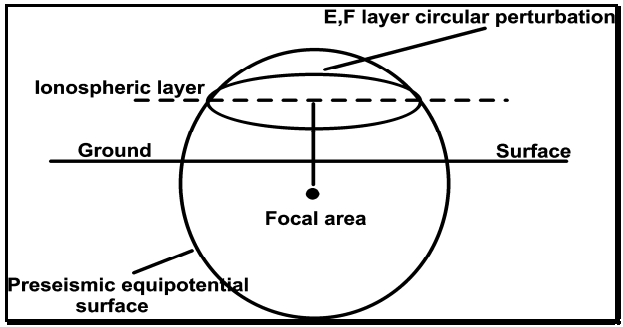

Fig. 3. A simplified model is presented of the perturbing of the ionospheric layers **E, F** mechanism, due to the presence of the preseismic, electric field, which is generated a few days before the occurrence of the pending, strong EQ.

A more detailed presentation of this mechanism is presented in the following figure **(4)**.

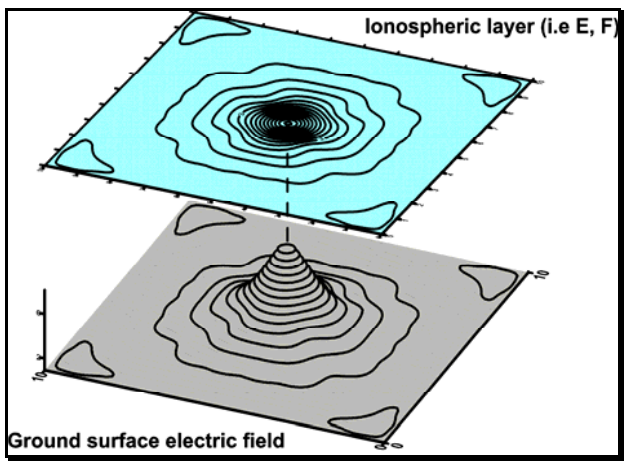

Fig. 4. Plasma – electron density perturbation of the ionospheric layers **E, F** (upper map), due to the presence of a preseismic, electric field on the ground surface (lower map), penetrating the ionospheric layers.

Actually, what happens is that, the concentric, equipotential lines, generated, by the intersection of the E, F ionospheric layers with the equipotential surfaces, which surfaces are generated from any triggered, physical mechanism in the focal area, form a local, circular, plane electrical field (its gradient pointing to its center) in the ionosphere, which attracts inwards or repels outwards, charged particles (ions), depending on the polarity of the electrical field. Therefore, in the ionosphere, circular form perturbations of electron – plasma density are generated. This is what is demonstrated in figure (**4**).

Although, this mechanism seems simple, other electrical fields which are generated in the ionosphere affect the created, preseismic, ionospheric circular perturbations. Consequently, the ionospheric data, must be demasked from such effects, before any conclusion is made about any suspected, epicenter area of a future strong EQ. Indicative examples, of such operations, have been presented by Depueva and Rotanova (2001).

The previously, presented, physical mechanism validates from another point of view (the one of ionospheric observations), the presence in the seismogenic area of an anomalous, preseismic, electrical field which gives rise to various, preseismic, electrical signals. This electrical field exhibits directional properties and therefore, it is possible to determine its location of origin by applying simple electric, potential theory physics laws.

In the present work the directional properties of the Earth's preseismic electric field will be demonstrated and its correlation to the corresponding same time period seismicity will be considered too.

## 2. Theoretical analysis.

The homogeneous ground Earth model which is adopted (Thanassoulas, 1991, 2007) implies that a point, current source, located in it, generates equipotential lines on the ground surface. In the case of a seismogenic area, where a strong earthquake will take place some time in the near future, the center of the generated, by the apparent point current source equipotential lines, coincides with the epicenter of the pending earthquake. Schematically it is shown in the following figure **(5)**.

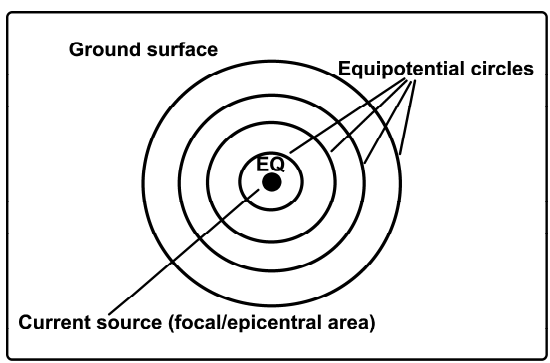

Fig. 5. Schematic presentation of the equipotential lines which are generated on ground surface by the apparent point, current source. This source is located in the focal area, which is centered at the epicentral area of the pending, strong earthquake.

At this stage, it is a simple task to calculate the azimuthal direction of the Earth's electric field intensity vector. The azimuthal direction of the electrical field intensity vector indicates the direction of the location of the apparent point current source, in relation to the registration site. Schematically, this procedure is presented in the following figure **(6)**.

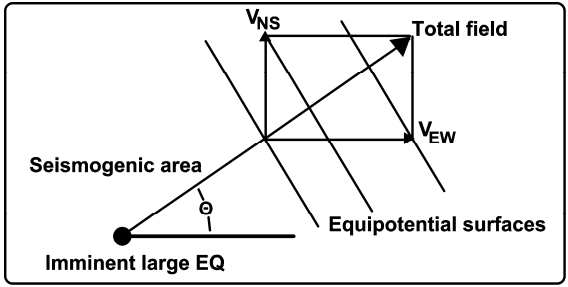

Fig. 6. Azimuthal direction **(θ)** of the Earth's electric field intensity vector calculation related to the location of the registration site.

The actual procedure is as follows:

The Earth's electric field is measured in two orthogonal components $V_{NS}$ and $V_{EW}$. The next step is to calculate the angle **(θ)**. This is achieved through the following trigonometric equation:

$$(\theta) = \arctan(V_{NS} / V_{EW}) \tag{1}$$

The equation **(1)** is valid assuming that, the registering system complies with the requirements of being a unit trigonometric circle. In other words, the electric dipoles which are used for the registration of the Earth's electric field, must be of the same length and moreover these must be oriented towards **N-S** and **E-W** directions, so that is achieved a geographical orientation of the Earth's electric field intensity vector. The registering dipole system must be equal to the trigonometric circle, shown in the following figure **(7)**.

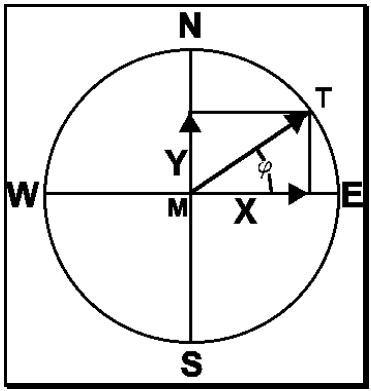

Fig. 7. The trigonometric unit circle is presented, which must reflect the registering system of the Earth's electric field, so that an azimuthal determination may be achieved of the direction of the electrical field intensity vector.

The dipoles, which are used for the registration of the Earth's electric field **(T)** **X** and **Y** components, must be oriented towards **E-W** and **N-S**. In practice, when a dipole system is to be located at a specific place, the most common situation, met, is that it is not possible to orient the dipoles exactly at **E-W** and **N-S** direction, but at more or less different angles. Therefore, "normalization" to **NS / EW** direction of the electric dipoles is required before any further processing of the electric data takes place.

During the overall seismic activation of a larger, seismic, prone area, it is understood that specific seismogenic areas may be activated in different periods of time. These seismogenic areas generate electrical signals which interfere with each other. Consequently, the registration of the Earth's electrical field corresponds to the total field which results from this interference.

The important fact is that, each time a focal area generates a strong signal. This signal prevails on all others and consist the main component of the electrical field to be analyzed. Therefore, its characteristic parameters will be mainly described in the analysis of the total electrical field.

**2.1. The theoretical model.**

Let us recall the equation:

$$\theta = \arctan(V_{NS}/V_{EW}) \tag{2}$$

Equation **(2)** will be studied, in particular, for the case of an oscillating electrical field. In the case of an oscillating current source the current takes the form of:

$$I = I_0 \sin(kt) \tag{3}$$

and therefore the equation that expresses the potential **V**, **(V = $I_0$R / 4πr)**, due to a current source **($I_0$)**, at a distance **(r)**, in a medium of resistivity **(R)**, is transformed into:

$$V = I_0 \sin(kt) \, r^{-1} \, R / (4*pi) \tag{4}$$

From equation **(4)** it is easily obtained that **V<sub>EW</sub>** and **V<sub>NS</sub>** take the form of:

$$V_{EW} = V_1 \sin(kt) \text{ and } V_{NS} = V_2 \sin(kt) \qquad (5)$$

where **V₁** and **V₂** are:

$$V_1 = I_0 \sin(kt)\cos(\theta) \, r^{-1} R / (4\pi) \qquad (6)$$

$$V_2 = I_0 \sin(kt)\sin(\theta) \, r^{-1} R / (4*\pi) \qquad (7)$$

From equations **(5)**, **(6)**, **(7)** results that:

$$\text{Angle } (\theta) = \arctan(V_2 / V_1) \qquad (8)$$

it is the same equation which is valid for the non-oscillating field. The scalar value **(Vt)** of the electrical field intensity vector at time **(t)** takes the form of:

$$Vt = \sin(kt)(V_1^2 + V_2^2)^{1/2} \qquad (9)$$

Thus, indicating that the intensity vector of the electric field oscillates along an axis **(fig. 8)** that forms an angle **(θ)** to the **EW** direction, calculated by the equation **(2).**

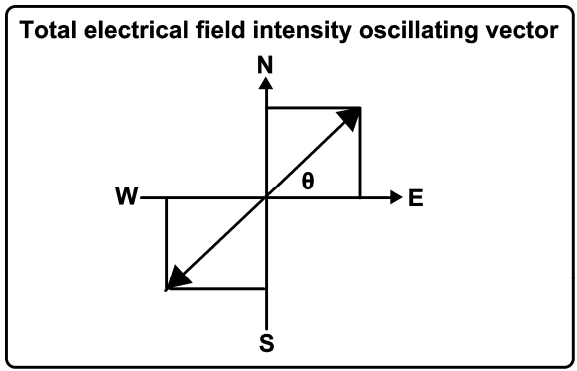

Fig. 8. **Vt** oscillates at an angle **(θ)** along the double black arrow.

Generally, when an external, oscillating, electrical field interferes with the initial one, the equation **(9)** takes the form of:

$$I = I_0 \sin(k_0)\sin[(k+k_1)t] \qquad (10)$$

Where: **(K₀)**, **(k)** and **(k₁)** are different angular velocities. As a result, angle **(θ)** is not anymore constant but becomes a function **(F)** of **(k₀, k, k₁),** as well as **(Vt)** is, too.

$$(\theta) = F(k_0, k, k_1) \text{ and } Vt = f(k_0, k, k_1) \qquad (11)$$

In this case **(Vt)** does not oscillate along a line but it generally prescribes an ellipse **(fig. 9).**

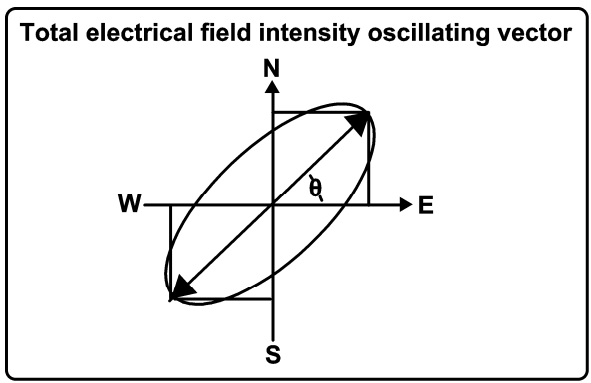

Fig. 9. **Vt** oscillates and prescribes an ellipse.

The following cases are representative examples of the later analysis.

### 2.2. Electrical field orthogonal components which were generated by different current sources.

In this case, the calculated angle (**θ**) varies in random since there is no correlation between the observed two components. The recorded, two signals (Mares, 1984) are shown in figure **(10)**. The corresponding, calculated electrical field intensity vectors are presented in figure **(11)**.

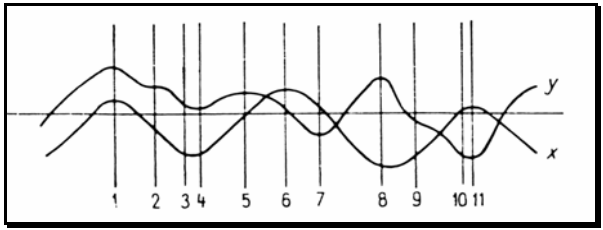

Fig. 10. Uncorrelated, orthogonal, electrical field components (after Mares, 1984).

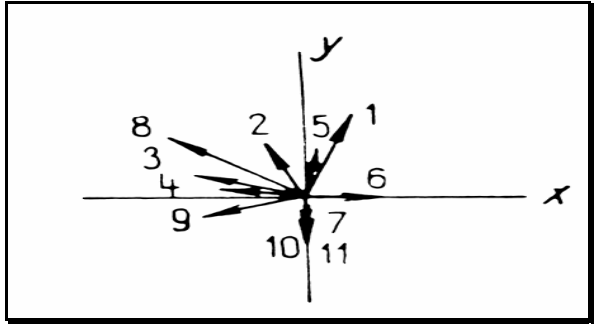

Fig. 11. Vectorgram azimuthal values calculated for the observed, uncorrelated, electrical, orthogonal components (after Mares, 1984).

### 2.3. Electrical field, orthogonal components, which were generated by the same current source.

In this case, the previous mathematical analysis is valid and can be applied on the obtained data. The two orthogonal components (Telford et al. 1976) are shown in figure **(12)**, while the corresponding polarization ellipse is presented in figure **(13)**.

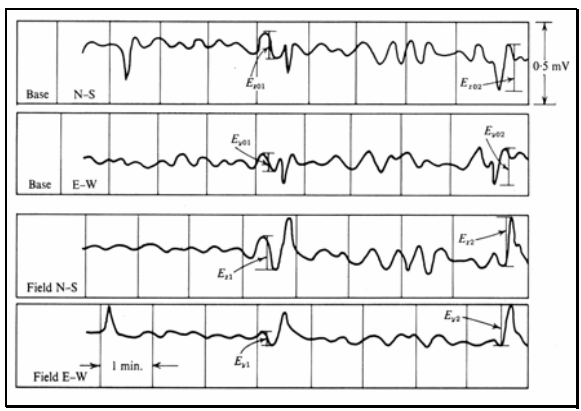 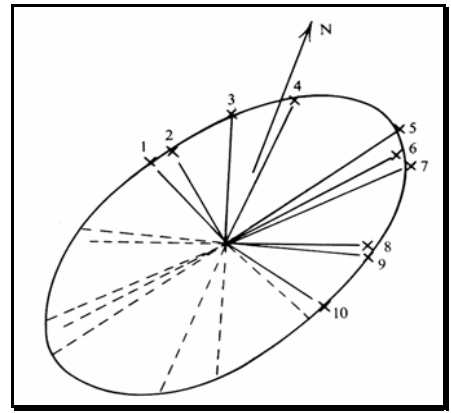

Fig. 12 - 13. **Left**: Orthogonal components of the electrical field observed, that corresponds to a single, current source. **Right**: Azimuthal values calculated, for the observed, correlated, electrical, orthogonal components (after Telford et al. 1976).

The analysis already presented requires that the data to be used must be free from any noise. This implies that some kind of low-pass filtering must be applied on the raw data beforehand an azimuthal calculation is made. This procedure is demonstrated by the following example.

Low-pass filtering was applied over data which correspond to a single anomaly of the electrical field, superimposed, over a larger, regional field (Patra and Mallick, 1980). The orthogonal components **(Ex)** and **(Ey),** are presented, in figure **(14).**

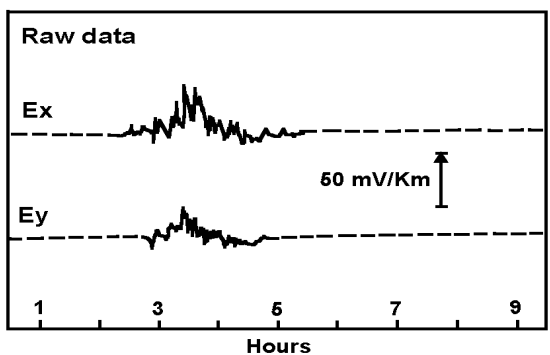

Fig. 14. Electrical field, orthogonal components (after Patra and Mallick, 1980) generated by the same, current source that corresponds to a local, electrical field anomaly. It is obvious, as it appears from the data, that high frequency noise has interfered with the original data. Therefore, the corresponding ellipse was calculated, **(a)** for the original data as it is, **(b)** after having applied some "low-pass" and **(c)** "band-pass" filtering. This operation is shown in figure **(15)**.

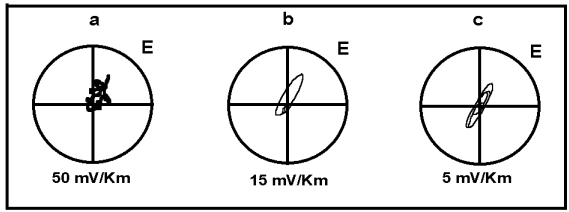

Fig. 15. Azimuthal values calculated, (after Patra and Mallick, 1980) for the observed, correlated, electrical, orthogonal components. Case (a) = raw data, (b) = 5$^{th}$-harmonic synthesis, (c) = band-pass filtered / $T_0$ = 10000 sec. 2.4. Electrical field orthogonal components which are generated by the interference of more than one current source.

Let us consider now the case of more than one current source, which affect the electrical field recorded by a monitoring site.

As an example, is studied the case of three current sources. In figure **(16)** is presented the location of three current sources **(EQ1, 2, 3)** in relation to the location of the monitoring site along with the expected, theoretical, equipotential lines.

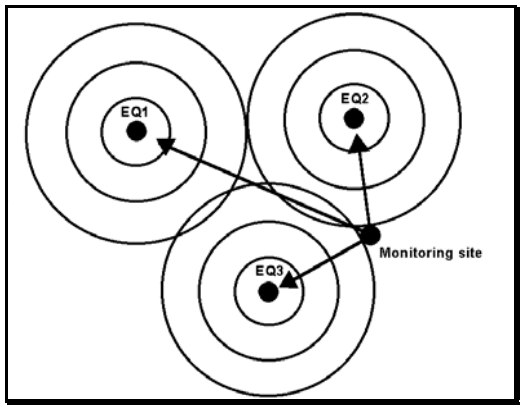

Fig. 16. Location of the **EQ1, 2, 3** current sources is presented, in respect to the location of the monitoring site.

An ellipse is generated **(fig. 17)** for each current source. Its major axis indicates the azimuthal direction of the Earth's oscillating, electric field, current source origin location, in relation to the location of the monitoring site.

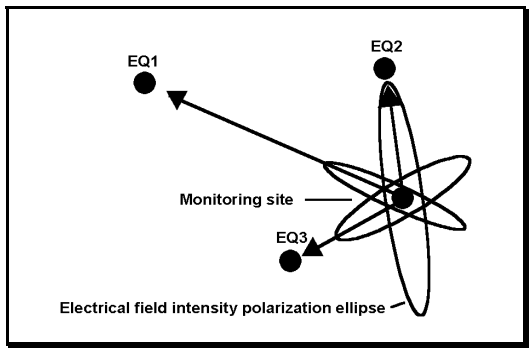

Fig. 17. Corresponding ellipses and electrical field intensity vector presentation for each current source **(EQ1, 2, 3)**.

If we take into account that the **EQ1, 2, 3** current sources do not evolve simultaneously, in time, it is made clear that a sequential, in time, registration of the orthogonal components of the electric field, will result in a successive peaking, towards the corresponding direction of the **(Vt)** value, and each time, a discrete, current source prevails, to form the registered total electrical field.

The later analysis was applied to the electrical signals which were observed during seismically active periods of time in the Greek territory. The epicentral (focal) areas which were activated seismically were considered as the current sources.

**The working hypothesis is as follows:**

**If the seismically, activated, regional focal areas emit such electrical signals, then the continuous recording of the two orthogonal components of the oscillating electrical field, at a monitoring site, must reveal the preferential, azimuthal directions of the intensity vectors, in relation to the monitoring site location. This will coincide with the corresponding areas which have already been activated seismically.**

**3. Real data analysis.**

The already presented analysis was applied on the data which were obtained by Volos **(VOL)** monitoring site (collaboration with Tsatsaragos, 2002). These data were analyzed with the same procedure by Thanassoulas and Klentos (2003). A brief presentation with actual examples follows. The two orthogonal components of the electric field were registered continuously, in a digital form, at a sampling interval of one minute.

The oscillating component of the electrical field is obtained by applying Fast Fourier Transform **(FFT)** and "band-pass" digital filtering to the recorded raw data. The center period of the band-pass filter is set at T = 24 hours.

A sample of raw data, as it is recorded, is shown in the following figure **(18)** while its filtered result is shown in figure **(19).**

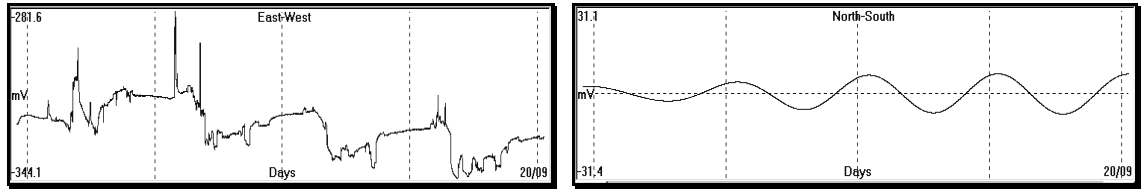

Fig. 18 - 19. Left: A sample of raw data, as it is recorded. Right: Filtered data are presented, as a result of "band-pass" filtering operation.

The already presented analysis of the Earth's electric field will be applied on a real data set. The obtained azimuthal directions will be compared to the concurrent seismicity for the same period of time.

**- Time span – recorded orthogonal components.**

The data which are used in this example extend from 18/06/2002 to 21/06/2002. A total recording of **4** days, whose oscillating component, of the electrical field registered, is presented in fig. **(20)**.

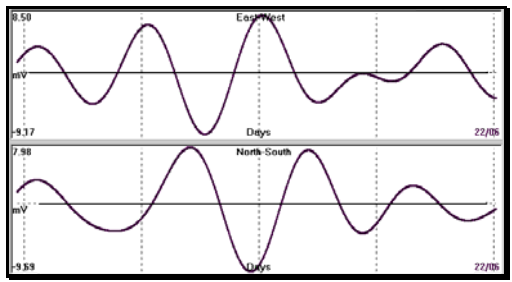

Fig. 20. Orthogonal, oscillating components, of the Earth's electric field are shown. The recording period extends from 18/06/2002 to 21/06/2002.

**- Polar diagram construction.**

The azimuthal direction of the electrical field intensity vector was calculated for the same time period at 1-minute intervals, and is presented in figure **(21)**. This calculation is performed by a specific software package constructed for this purpose. Its main feature is the detection of the maximum values of the calculated intensity vector, along the time axis of the recording, and the location, in a polar diagram, of its azimuthal direction. Each time a maximum value was identified the corresponding vector color changed, so that it facilitated the visual follow-up of this processing. The detected, successive peaks indicate the azimuthal directions of the consecutive, seismically, activated areas.

Finally, the azimuthal directions which were calculated from the maxims of the electrical field intensity vectors are superimposed **(fig. 22)** on the map of the location of the corresponding earthquakes that occurred during this period of time.

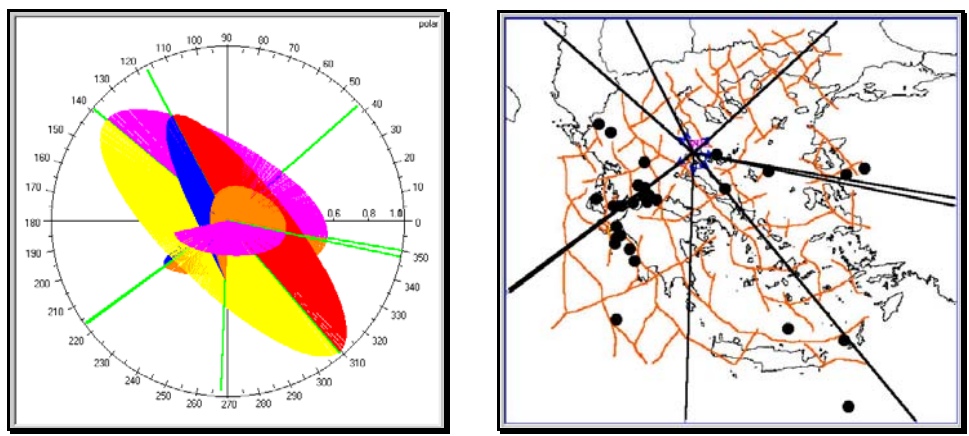

Fig. 21 - 22. **Left**: Polar diagram of the azimuthal direction changes of the calculated, electrical field intensity vector. **Right**: Calculated azimuthal directions (solid black lines) compared with the location of the earthquakes (**Ms>3.0R**) along the same registration period of time.

It is obvious that the majority of the EQs are located along the calculated already azimuthal directions. The observed, discrepancy for some of the earthquakes or azimuthal directions will be discussed later on.

### 3.1. Other examples.

Some more examples, that foster the validity of the methodology, follow. The figures are presented in the same order as: oscillating signal, corresponding, polar diagram and correlation map of azimuthal directions with EQs location. At each case, the low-level threshold magnitude of the compared EQs is presented, too.

#### **Example – 1**

Time period: 2002/03/16-18
Detected EQs low magnitude threshold: **Ms = 4.0R**

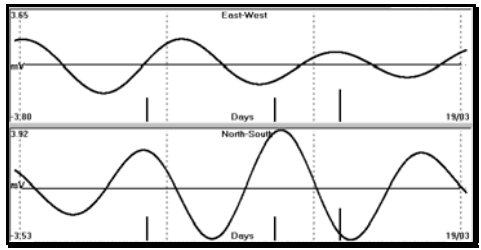

Fig. 23. Signal recorded, vertical black lines indicate concurrent seismicity.

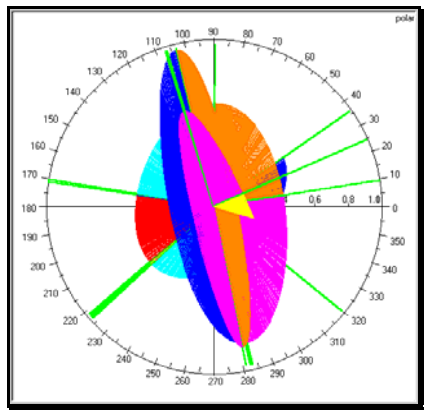
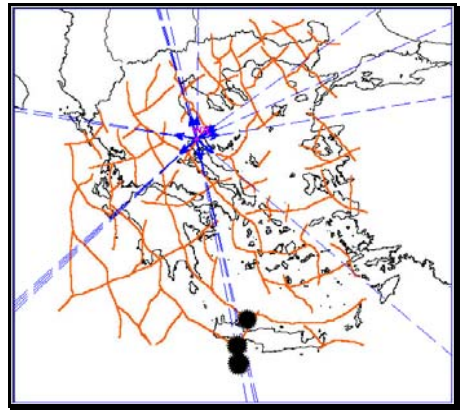

Fig. 24 - 25. **Left**: Corresponding, polar diagram. **Right**: Correlation map of azimuthal directions to EQs location (solid black circles).

#### **Example – 2**

Time period: 2002/04/18-21
Detected EQs magnitude-low threshold: **Ms = 4.5R**

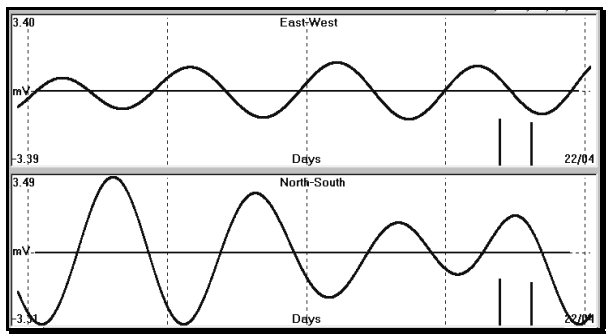

Fig. 26. Signal recorded, vertical black lines indicate concurrent seismicity.

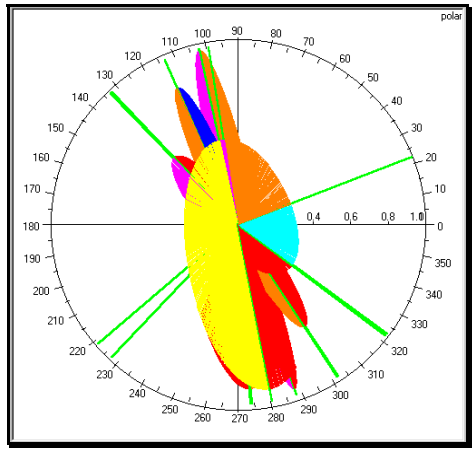 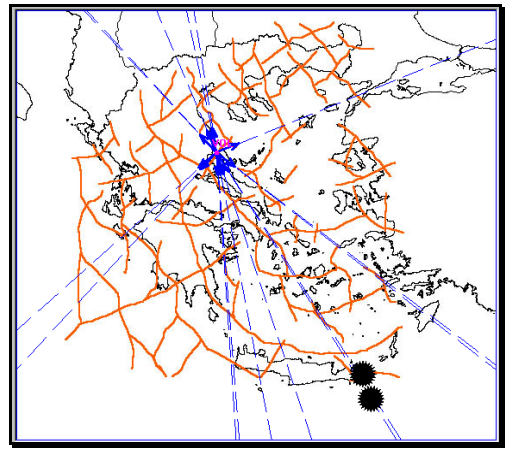

Fig. 27 – 28. **Left**: Corresponding, polar diagram. **Right**: Correlation map of azimuthal directions to EQs location.

**Example – 3**

Time period: 2002/07/23-26
Detected EQs magnitude-low threshold: **Ms = 3.0R**

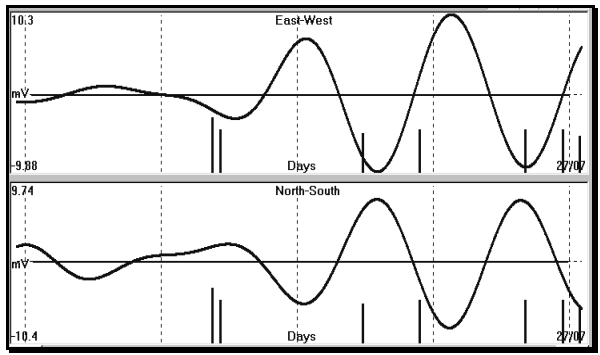

Fig. 29. Signal recorded, vertical black lines indicate concurrent seismicity.

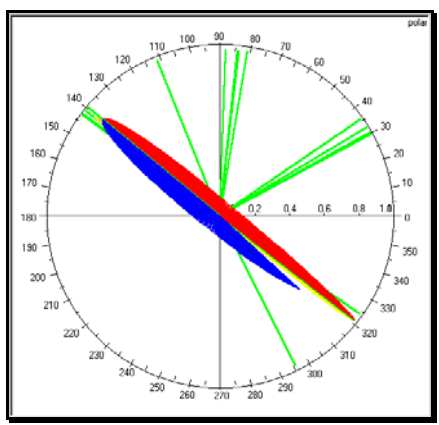 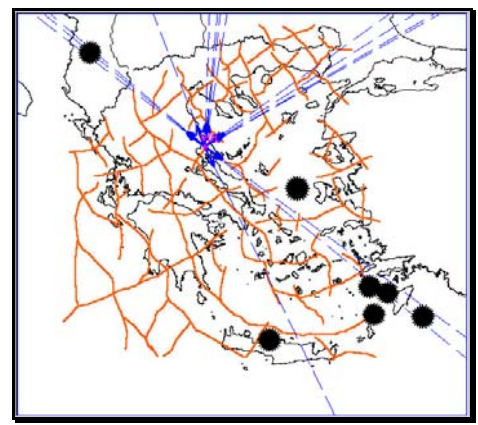

Fig. 30 - 31. **Left**: Corresponding, polar diagram. **Right**: Correlation map of azimuthal directions to EQs location.

**Example – 4**

Time period: 2002/09/03-05
Detected EQs magnitude-low threshold: **Ms = 4.0R**

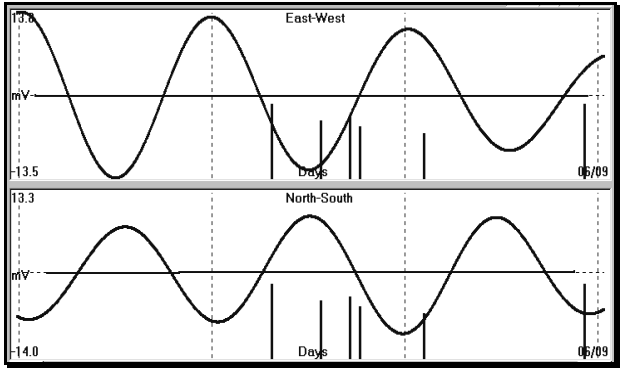

Fig. 32. Signal recorded, vertical black lines indicate concurrent seismicity.

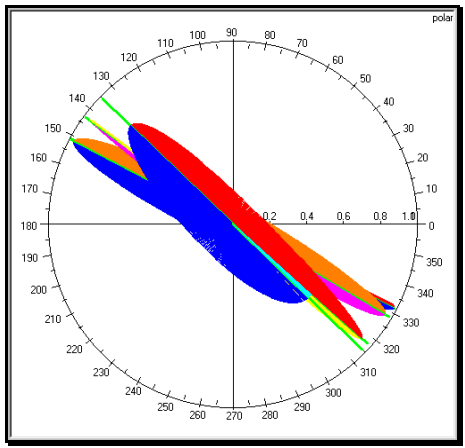
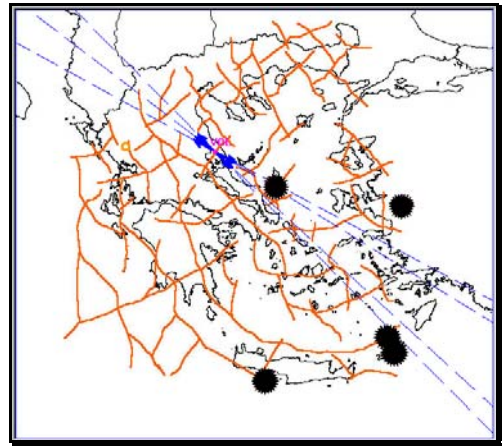

Fig. 33 - 34. **Left**: Corresponding, polar diagram. **Right**: Correlation map of azimuthal directions to EQs location.

**3.2. The recent example from Methoni seismogenic area (**Main event on February 14th, 2008, Ms = 6.7R**)**

So far, the correlation of the azimuthal directions, of the major ellipses axis of the intensity vector of the oscillating electric field, generated by the triggered mechanisms, to the epicentral areas of the corresponding earthquakes has been presented. The example to follow is a quite different one. The already presented property of the Earth's electric preseismic field will be used in **a predictive mode** for the location of the epicenter of a future large earthquake (Thanassoulas, 2008a, b).

In this example, after the occurrence of the Methoni EQ, the oscillating Earth's electric field was analyzed at three different monitoring sites (**PYR, ATH, HIO**). The triangulation of the determined three azimuthal directions indicated the future epicentral area of a future large EQ. The convergence of the determined oscillating vectors is presented by concentric green circles in figure (**35**).

During the next 50 days after the analysis of the Earth's electric field (21-22/2/2008), a quite large number (**11**) of EQs of Ms>=5.0R did take place in the already predicted epicentral area (fig. **36**).

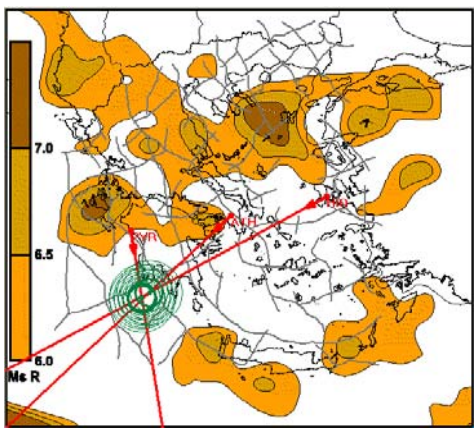
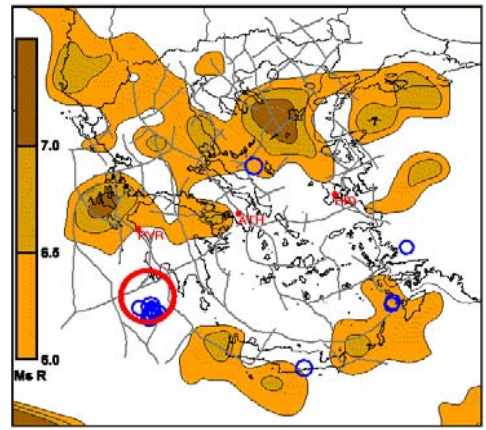

Fig. 35 – 36. **Left**: epicentral area (concentric green circles) determined by the triangulation of the azimuthal direction of the Earth's oscillating electric preseismic field. **Right**: EQs with Ms>=5.0R (blue circles) which occurred within the time period of 20/2 – 10/4/2008. The EQs which occurred in the predicted (red circle) epicentral area are 11 out of a total of 15 EQs (Thanassoulas, 2008a, b).

## 4. Discussion - Conclusions

So far, the calculated, by one monitoring site, electrical field intensity vectors are distinguished into two groups.

**The first one**, generally, does not correlate with any known azimuthal direction, in relation to the monitoring site, of the EQs that occurred during the period of time of the study.

This group of vectors is clustered in narrow azimuthal bands, in contrast to an expected, normal, random, azimuthal direction distribution which reflects a random process. The later, probably indicates that, some seismogenic areas were excessively stress-loaded but didn't end up to the occurrence of any EQ. Another explanation is the following: since the electric field oscillates at an angle (**φ**) then a second azimuth is observed at an angle (**φ + π**). These azimuth directions are easily detected on the polar graph.

**The second group** of vectors, in all the presented cases, exhibits a very good fit of azimuthal directions of calculated vectors and EQs. Comparing this group with the various magnitude levels of the EQs, it correlates very well, some times even down to magnitudes of M = 3.0R.

The later shows not only the high sensitivity of the used method, but indicates very well that the generated, electrical field, at the focal area, can be detected at large distances from it. This can be explained by the highly resistant crustal seismogenic layer where the electrical dissipation of the transmitted electrical energy is either negligible or very small.

In conclusion, this analysis shows that, due to the applied stress load, the seismically, activated areas, generate electrical signals that can be used for the calculation of their (seismically activated areas) azimuthal direction, **in relation to a single monitoring site**. The analyzed data and the presented examples indicate that the later can be utilized even at low level preseismic electric signals generated by EQs of relatively low magnitudes.

This methodology opens up the possibility for the "seismic electrical potential status" of a largely extended seismogenic region to be monitored **"a priori" in real time, even if there is no seismological, preseismic evidence at all,** by installing a network of monitoring sites, evenly distributed, over a large, seismogenic area (i.e. Greece),

The use of the directional property of the oscillating Earth's electric field, calculated **at three different monitoring sites**, provides us with the possibility to use triangulation for the estimation of the epicentral area of a future large EQ. The later was validated in the case of the Methoni, Greece seismogenic area.

## 5. References.

Davies, K., and Baker, D., 1965. Ionospheric effects observed around the time of the Alaskan earthquake of March 28 1964, J. Geophys. Res., 70, pp. 2251-2253.
Depueva, A., and Rotanova, N., 2001. Low-latitude ionospheric disturbances associated with earthquakes., Annali di Geofizika, 44, pp. 221-228.
Di Giovambattista, R., Tyupkin, Y., 2004. Seismicity patterns before the M=5.8 2002, Palermo (Italy) earthquake: seismic quiescence and accelerating seismicity., Tectonophysics, 384, pp. 243-255.
Horn, M., Boudjada, M., Biernat, H., Lammer, H., Schwingenschuh, K., Prattes, G., 2007. Model calculation of the electrostatic field penetration into the ionosphere., Geophysical Researc Abstracts, Vol. 9, 06582, European Geosciences Union.
Ifantis, A., Tselentis, G., Varotsos, P., and Thanassoulas, C., 1993. Long-term variations of the earth's electric gield preceding two earthquakes in Greece., Acta Geophysica Polonica, Vol. XLI, no.4, pp.337-350.
Keilis-Borok, V.I and Rotwain, I.M., 1990. Diagnosis of time of increased probability of strong earthquakes in different regions of the world: algorithm CN., Physics of the Earth and Planetary Interiors, 61, pp. 57-72.
Mares, S., 1984. Introduction to Applied Geophysics., Reidel Publishing Company, Dordrecht / Boston / Lancaster.
Mizutani, H., Ishido, T., Yokohura, T., Ohnishi, S., 1976. Electrokinetic phenomena associated with earthquakes., Geophysical Research Letters, Vol. 3, No. 7., pp. 365-368.
Moore, W., 1964. Magnetic disturbances preceding the 1964 Alaska Earthquake., Nature, 203., pp. 508-512.
Ogata, Y., Zhuang, J., 2006. Space-time ETAS models and an improved extension., Tectonophysics, 413, pp. 13-23
Patra, H.P., and Mallick, K., 1980. Geosounding Principles, 2., Time-varying Geoelectric Soundings., Elsevier, Amsterdam/Oxford/New York.
Pham, V., Boyer, D., Perrier, F., Le Mouel, J., 2001. Generation mechanisms of telluric noises in ULF band: possible sources for the so-called "seismic electric signals' (SES)., C.R. Acad. Sci. Paris, Sciences de la Terre et de planets/ Earth and Planetary Sciences, 333, pp. 255-262.
Pulinets, S., 2004. Ionospheric precursors of earthquakes; Recent advances in theory and practical applications., TAO, Vol. 15, No. 3, pp. 413-435.
Pulinets, S., 2006. Space technologies for short-term earthquake warning., Advances in Space Research, 37, No. 4, pp. 643-652.
Pulinets, S., Legen'Ka, A., Gaivoronskaya, T., Depuev, V., 2003. Main phenomenological features of ionospheric precursors of strong earthquakes., Journal of Atmospheric and Solar-Terrestrial Physics., 65, pp. 1337-1347.
Pulinets, S., Gaivoronska, T., Contreras, L., and Ciraolo, L., 2004. Correlation analysis technique revealing ionospheric precursors of earthquakes., Natural Hazards and Earth System Sciences., 4, pp. 697-702.
Romachkova, L., Kossobokov, V., Panza, G., Costa, G., 1998. Intermediate-term predictions of earthquakes in Italy: Algorithm M8., Pure and Applied Geophysics., 152, pp. 37-55.
Sobolev, G., 2001. The examples of earthquake preparation in Kamchatka and Japan., Tectonophysics, 338, 269-279.
Sobolev, G., Huang, Q., Nagao, T., 2002. Phases of earthquake's preparation and by chance test of seismic quiescence anomaly., Journal of Geodynamics, 33, pp. 413-424.
Telford, W., Geldart, L., Sheriff, R., Keys, D., 1976. Applied Geophysics, Cambridge University Press, London, UK.
Thanassoulas, C., 1991. Determination of the epicentral area of three earth-quakes (Ms>6) in Greece, based on electrotelluric currents recorded by the VAN network., Presented in: EUG – VI meeting, Strasbourg, France.
Thanassoulas, C., 1991a. Determination of the epicentral area of three earth-quakes (Ms>6) in Greece, based on electrotelluric currents recorded by the VAN network., Acta Geoph. Polonica, Vol. XXXIX, no. 4, pp. 273 – 287.
Thanassoulas, C., 2007. Short-term Earthquake Prediction, H. Dounias & Co, Athens, Greece. ISBN No: 978-960-930268-5

Thanassoulas, C., 2008a. The electric field of the Earth after the occurrence of the February 14th, 2008, Ms = 6.7R EQ in Greece. Its implications towards the prediction of a probable future large EQ., arXiv.org/0802.3752 [physics.geo-ph].

Thanassoulas, C., 2008b. Seismicity observed, at Methoni seismogenic area, Greece, after the analysis of the recorded Earth's electric field of 21/2/2008 – 22/2/2008, at PYR, ATH and HIO monitoring sites, Greece., arXiv.org/0805.2125 [physics.geo-ph].

Thanassoulas, C., Klentos, V., 2003. Multidirectional analysis of low-level 34 hours period oscillatory earthquake precursory electrical signals., IGME, Open File Report A. 4401, Athens, Greece, pp. 1-27.

Varotsos, P., Alexopoulos, K., Nomicos, K., Papaioannou, G., Varotsou, M., Revelioti-Dologlou, E., 1981. Determination of the epicenter of impending earthquakes from precursor changes of the telluric current., Proceedings of Greek Academy of Sciences, 56, pp. 434-446.

Varotsos, P., and Alexopoulos, K., 1984. Physical properties of the variations of the electric field of the earth preceding earthquakes. II. Determination of epicenter and magnitude., Tectonophysics, 110, pp. 99-125.

Wyss, M., and Baer, M., 1981. Earthquake Hazard in the Hellenic Arc., in: Earthquake Prediction, An International Review, American Geophysical Union, Washington DC., USA.